\documentclass[twocolumn,showpacs,preprintnumbers,amsmath,amssymb]{revtex4}
\usepackage{graphicx}% Include figure files
\usepackage{dcolumn}% Align table columns on decimal point
\usepackage{bm}% bold math

\begin{document}

\title{Space-time sensors using multiple-wave atom levitation}
\author{F. Impens$^{1,2}$ and Ch. J. Bord\'{e}$^{1,3}$}

\affiliation{$^{1}$ SYRTE, Observatoire de Paris, 61 avenue de
l'Observatoire, 75014 Paris, France}

\affiliation{$^{2}$ Instituto de Fisica, Universidade Federal do
Rio de Janeiro. Caixa Postal 68528, 21941-972 Rio de Janeiro, RJ,
Brasil}

\affiliation{$^{3}$ Laboratoire de Physique des Lasers, Institut
Galil\'{e}e,  F-93430 Villetaneuse, France}

\date{\today }

\begin{abstract}
The best clocks to date control the atomic motion by trapping the
sample in an optical lattice and then interrogate the atomic
transition by shining on these atoms a distinct laser of
controlled frequency. In order to perform both tasks
simultaneously and with the same laser field, we propose to use
instead the levitation of a Bose-Einstein condensate through
multiple-wave atomic interferences. The levitating condensate
experiences a coherent localization in momentum and a controlled
diffusion in altitude. The sample levitation is bound to resonance
conditions used either for frequency or for acceleration
measurements. The chosen vertical geometry solves the limitations
imposed by the sample free fall in previous optical clocks using
also atomic interferences. This configuration yields multiple-wave
interferences enabling levitation and enhancing the measurement
sensitivity. This setup, analogous to an atomic resonator in
momentum space, constitutes an attractive alternative to existing
atomic clocks and gravimeters.
\end{abstract}


\maketitle

\begin{figure}[htbp]
\begin{center}
\includegraphics{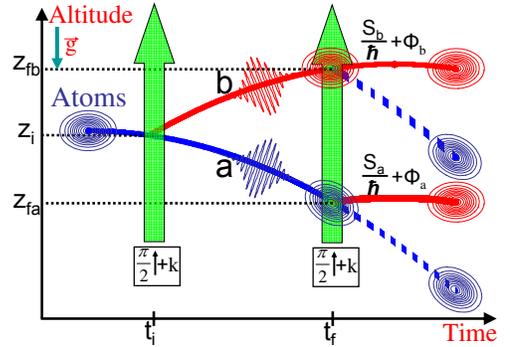}
\end{center}
\caption{Action of a pair of copropagating $\pi/2$-pulses on a
free-falling atomic wave-packet. The phase difference between the
two outgoing wave-packets comes from the classical action and
laser phase acquired on each path, and from the distance of their
centers.} \label{fig:sequence deux pisurdeux}
\end{figure}

The light-matter interaction enables the exchange of momentum
between an electromagnetic field and individual atoms: each atom
emitting or absorbing a photon experiences simultaneously a change
of internal level and a recoil reflecting momentum conservation.
This well-controlled momentum transfer can be used to engineer
correlations between the motional and the internal atomic state.
This is the principle underlying Bord\'e-Ramsey atom
interferometers~\cite{Borde84,BordeAtomInt89,Berman}, which are the
building blocks of our system. Such interferometers consist in the
illumination of moving two-level atoms with a first pair of light
pulses separated temporally and propagating in the same direction
[Fig.\ref{fig:sequence deux pisurdeux}], followed by a second pair
of pulses coming from the opposite direction. Each pulse operates a
$\pi/2$-rotation on the vector representing the atomic density
matrix on the Bloch sphere: applied on a given internal state, it
acts as an atomic beamsplitter by creating a quantum superposition
of two atomic states with distinct internal levels and momenta.
Horizontal Bord\'e-Ramsey interferometers have been used to build
optical clocks~\cite{Wilpers07}. This system presents, however, two
drawbacks: the free fall of the atoms through the transverse lasers
probing their transition limits the interrogation time and induces
undesirable frequency shifts~\cite{Trebst01}. This led the metrology
community to privilege atomic clocks~\cite{BestOpticalClocks} built
around atomic traps~\cite{Katori03}, able to control the sample
position. Such systems have become sufficiently accurate to probe
fundamental constants~\cite{FundamentalTests}. We propose instead to
circumvent these limitations with a multiple-wave atom
interferometer~\cite{MultipleWaveAtomInterferometry} in levitation,
which comprises a succession of vertical Bord\'e-Ramsey atom
interferometers. This strategy combines the best aspects of optical
clocks based on atom traps and on atom interferometers: it prevents
the sample free fall without using optical potentials likely to
cause spurious frequency shifts. The recent experimental
achievement~\cite{Hughes09} of a sustainable levitation of coherent
atomic waves with synchronized light pulses~\cite{Impens06} strongly
supports the feasibility of this method.

Our purpose is to provide a controlled vertical momentum transfer
to the atoms, eventually enabling their levitation, through the
repetition of a four vertical $\pi/2$-pulse sequence. Momentum
kicks are achieved by performing two successive population
transfers with vertical pulses propagating in opposite directions:
starting from the adequate atomic state, one obtains successively
the absorption of an upward photon followed by the emission of a
downward one, imparting a net upward momentum to the atoms. This
leads us to consider the point illustrated in
Fig.~\ref{fig:sequence deux pisurdeux}: when do two time-separated
$\pi/2$-pulses realize a full population transfer?
To achieve this, one must indeed compensate the phase induced by
the external atomic motion in the time-interval through fine-tuned
laser phases. This phase adjustment is at the heart of our
proposal, since it provides the resonance condition serving for
the laser frequency stabilization in the clock operation. The
$\pi/2$-pulse sequence then yields a \textit{conditional} momentum
transfer, controlled by this resonance condition, which
distinguishes this process from atomic Bloch
oscillations~\cite{BlochSystems}. Long $\pi$-pulses, realizing
conditional population transfers~\cite{Berman}, could also perform
such levitation~\cite{Impens06,Hughes09}. Yet there are several
benefits in privileging $\pi/2$-pulses: the atomic illumination
time is drastically reduced, the pulses address a broader
distribution of atomic momenta, and a better sensitivity is
obtained through a wider interferometric area~\cite{Berman}. To
obtain the resonance condition, we consider a dilute sample of
two-level atoms evolving in the gravity field -taken as uniform-
according to the Hamiltonian $H= \frac {p^2} {2 m}+ m g z$.
 It is initially in the lower state $a$ and described by the Gaussian
wave-function~\cite{footnote1} $\psi_a(\mathbf{r},t_0)=
 \frac {\pi^{-3/2}} {\sqrt{w_{x0} w_{y0} w_{z0} }}
e^{-\frac {1} {2} \left( \frac {(x-x_{i})^2} {w_{x0}^{2}}+\frac
{(y-y_{i})^2} {w_{y0}^{2}}+\frac {(z-z_{i})^2} {w_{z0}^{2}} \right)+
\frac {i} {\hbar} \mathbf{p}_{ia} (t_0) \cdot
(\mathbf{r}-\mathbf{r}_{i})}$. After two $\pi/2$-pulses, performed
at the times $t_i$ and $t_f=t_i+T$, the initial wave-packet has been
split into four packets following two possible intermediate
trajectories. The excited state wave-function receives two
wave-packet contributions coming from either path and associated
with the absorption of a photon at times $t_i$ and $t_f$, of common
central momentum $\mathbf{p}_{f }=\mathbf{p}_i+\hbar \mathbf{k}+ m
\mathbf{g} T$ and respective central positions $\mathbf{r}_{f a}$
and $\mathbf{r}_{f b}$. These wave-packets acquire a
 phase $S_{a,b}/\hbar$ reflecting the action on
each path~\cite{BordeMetrologia2002} and a laser phase
$\phi_{a,b}$ evaluated at their center for the corresponding
interaction time. Both contributions to the excited state are
phase-matched if the following relation~\cite{footnote2} is
verified
\begin{equation}
\label{eq:basic condition} - \mathbf{p}_f \cdot \mathbf{r}_{f, a}+
S_{a}+ \hbar \phi_{a}= - \mathbf{p}_f \cdot \mathbf{r}_{f, b}+
S_{b}+ \hbar \phi_{b} \, .
\end{equation}
The terms $\mathbf{p}_f \cdot \mathbf{r}_{f a,b}$ reflect the
atom-optical path difference between both wave-packets at their
respective centers. The central time $t_{c}=(t_i+t_f)/2$ is used
as phase reference for the two successive pulses. The phases
$\phi_{b}$ and $\phi_{a}$, provided respectively by the first and
the second $\pi/2$-pulse, read $\phi_{b,a}= \mathbf{k} \cdot
\mathbf{r}_{i,f a} - \omega_{1,2} (t_{i,f}-t_{c})+\phi^0_{1,2}$.
The action is given by $S_{a,b}=  m g^2 T^3/3+ \mathbf{p}_{i a, i
b} \cdot \mathbf{g} T^2+( p_{i a,i b}^2/2m - m g z_i-E_{a,b} ) T$.
Condition~\eqref{eq:basic condition} is fulfilled if the
frequencies $\omega_{1,2}$ of the first and second pulses
 are set to their resonant values
\begin{equation}
\label{eq:resonance frequencies} \omega_{1,2} \: = \: \frac {1}
{\hbar} \: \left( E_b+ \frac {(\mathbf{p}_{1,2}+ \hbar
\mathbf{k})^2} {2 m} - E_a - \frac {p_{1,2}^2} {2 m} \right)
\end{equation}
with $\mathbf{p}_{1}= \mathbf{p}_{i}$ and
$\mathbf{p}_{2}=\mathbf{p}_{i} + m \mathbf{g} T$, and if the
constant phases $\phi^{0}_{1,2}$ satisfy $\phi^{0}_{1}=
\phi^{0}_{2}$. If these conditions are fulfilled, and if the
sample coherence length $w$ is much larger than the final
wave-packet separation $|\mathbf{r}_{f, a}- \mathbf{r}_{f, b}|$ -
or equivalently if the Doppler width $k \Delta p  /m$ experienced
by the travelling atoms is much smaller than the frequency $1/T$
-, one obtains an almost fully constructive interference in the
excited state. The succession of two $\pi/2$-pulses then mimics
very efficiently a single $\pi$-pulse, the quantum channel to the
lower state being shut off by destructive interferences.  A key
point is that condition~\eqref{eq:basic condition}, expressing the
equality of the quantity $I=- \mathbf{p}_f \cdot \mathbf{r}_{f}+
S+ \hbar \phi$ for both paths, is independent of the initial
wave-packet position. This property allows one to address
simultaneously the numerous wave-packets generated in the $\pi/2$
pulse sequence.

\begin{figure}[htbp]
\begin{center}
\includegraphics{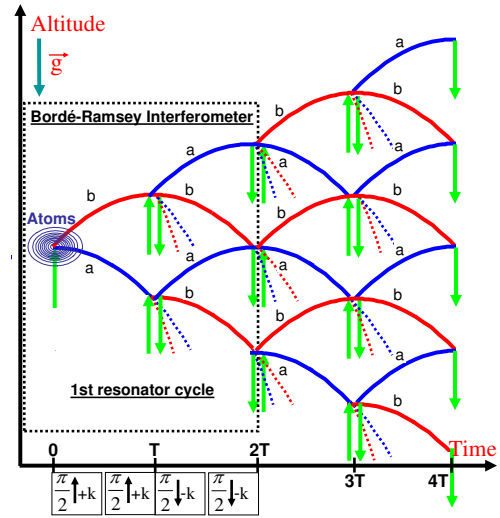}
\end{center}
\caption{(Color online)Levitating atomic trajectories in the
sequence of pulses. The first four pulses generate a vertical
Bord\'e-Ramsey interferometer. The central positions of the
wave-packets explore a network of paths which doubles at each
laser pulse.} \label{fig:arches}
\end{figure}

Applying a second sequence of two $\pi/2$ pulses with downward
wave-vectors~\cite{footnote3}, one obtains a vertical
Bord\'e-Ramsey interferometer bent by the gravity field sketched
on Fig.~\ref{fig:arches}.
 Starting with a sufficiently coherent sample in the lower state, and with
well adjusted frequencies~\eqref{eq:resonance frequencies}  and
ramp slopes, the previous discussion shows that a net momentum
transfer of $2 \hbar \mathbf{k}$ is provided to each atom during
the interferometric sequence. For the special interpulse duration
\begin{equation}
\label{eq:resonance T}
 T:=T^{0}= \frac {\hbar k} {mg},
\end{equation}
 these atoms end up in the lower
state with their initial momentum. Two major benefits are then
expected. First, the periodicity of the sample motion in momentum
gives rise to levitation. Second, only two frequencies, given
by~\eqref{eq:resonance frequencies}, are involved in the
successive resonant pairs of $\pi/2$ pulses. In particular, the
first and the fourth pulse of the Bord\'e Ramsey interferometer,
as well as the second and the third one, correspond to identical
resonant frequencies: $\omega^0_{1}=\omega^0_{4}$ and
$\omega^0_{2}=\omega^0_{3}$.

If the previous conditions are fulfilled, the repetition of the
interferometer sequence gives rise to a network of levitating
paths - sketched on Fig.~\ref{fig:arches} - reflecting the
diffusion of the atomic wave in the successive light pulses. The
same laser field is used to levitate the sample and to perform its
interrogation, generating a clock signal based on either one of
the two frequencies $\omega^0_{1},\omega^0_{2}$. Our measurement
indeed rests on the double condition (\ref{eq:resonance
frequencies}, \ref{eq:resonance T}), which must be fulfilled to
ensure this periodic motion: should the parameters
$(T,\omega_{1,2,3,4})$ differ from their resonant values
$(T^0,\omega^0_{1,2,3,4})$, the outgoing channels would open again
and induce losses in the levitating cloud, which can be tracked by
a population measurement. We expect multiple wave interference to
induce a narrowing of the resonance curve associated with the
levitating population around this condition.

  We have investigated this conjecture through a numerical simulation.
 The considered free-falling sample is taken at zero
temperature, sufficiently diluted to render interaction effects
negligible, and described initially by a macroscopic Gaussian
wave-function. Its propagation in-between the pulses is obtained by
evaluating a few parameters: central position and momentum following
classical dynamics, widths satisfying $w_{x,y,z}^{2}(t)= w_{x,y,z
0}^{2}+ \frac {\hbar^2} {4 m^2 w_{x,y,z 0}^{ 2} } (t-t_0)^2$, and a
global phase proportional to the action on the classical
path~\cite{BordeMetrologia2002}. Interaction effects may be
accounted for perturbatively with a generalized ABCD matrix
propagation formalism~\cite{ImpensABCD09}. The diffusion of atomic
packets on the short light pulses is efficiently modeled by a
position-dependent Rabi matrix~\cite{Berman} evaluated at the packet
center. While the evolution of each wave-packet is very simple,
their number -doubling at each light pulse- makes their book-keeping
a computational challenge. This difficulty, intrinsic to the
classical simulation of an entangled quantum state, has limited our
investigation to a sequence of sixteen pulses, involving $28$
levitating Bord\'e-Ramsey interferometers associated with the
resonant paths. The number of atomic waves involved ($N \simeq
64000$) is nonetheless sufficient to probe multiple-wave
interference effects.

The atomic transition used in this setup should have level lifetimes
longer than the typical interferometer duration (ms). Possible
candidates are the Ca, the Sr, the Yb, and the Hg atoms, which have
a narrow clock transition in their internal structure. These atoms
should be cooled at a temperature in the nano-Kelvin range,
preferably in a vertical cigar-shaped condensate, in order to
guarantee a sufficient overlap of the interfering wave-packets and
preserve a significant levitating atomic population. We consider a
cloud of coherence length $w=100 \: \mu m$ much larger than the
wave-packets separation $2 h \simeq 15 \:\mu m$. Fig.~\ref{fig:clock
results} shows the levitating and the falling atomic population in
the lower state as a function of the frequency shift $\delta \omega$
from the resonant frequencies $\omega^0_{1,2,3,4}$. It reveals a
fully constructive interference in the levitating arches when
resonance conditions are fulfilled, as well as the expected
narrowing of the central fringe associated with the levitating
wave-packets. Falling wave-packets yield secondary fringe patterns
with shifted resonant frequencies, which induce an asymmetry in the
central fringe if the total lower state population is monitored.
This effect, critical for a clock operation, can nonetheless be
efficiently attenuated by limiting the detection zone to the
vicinity of the levitating arches. This strategy improves as the
levitation time increases: the main contribution to the ``falling''
background comes then from atoms with a greater downward momentum
and thus further away from the detection zone. Besides, multiple
wave interferences sharpen the symmetric ``levitating'' central
fringe fast enough to limit the effect of the asymmetric background
of falling fringes. Considering a shift $\delta T$ from the resonant
duration $T_0$, one obtains also a central fringe narrowing as the
number of pulses increases and thus an improved determination of
acceleration $g$ through condition~\eqref{eq:resonance T}
~\cite{Impens06,Hughes09}. Multiple wave interferences thus improve
the setup sensitivity in both the inertial and frequency domains.

To keep the sample within the laser beam diameter, it is necessary
to use a transverse confinement, which may be obtained by using
laser waves of spherical wave-front for the
pulses~\cite{Impens06}. In contrast to former horizontal
clocks~\cite{Wilpers07}, the atomic motion is here collinear to
the light beam, which reduces the frequency shift resulting from
the wave-front curvature. A weak-field treatment, to be published
elsewhere, shows that this shift is proportional to the ratio $
\Delta \omega_{curv.} \propto k \langle v_{\bot}^2 \rangle T /R$,
 involving the average square transverse velocity
$\langle v_{\bot}^2 \rangle$ and the field radius of curvature $R$
at the average altitude of the levitating cloud. Let us note that
our proposal implies technological issues which must be solved to
achieve accurate measurements, but they are no more challenging
than those of current atomic clocks and sensors. The final
population in a given internal state can be monitored by using a
time-of-flight absorption imaging with a resonant horizontal laser
probe~\cite{Hughes09}.

\begin{figure}[htbp]
\begin{center}
\includegraphics{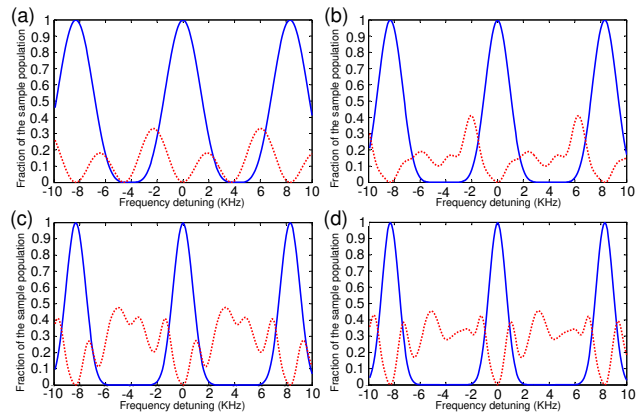}
\end{center}
\caption{(Color online)Fraction of the total sample population
levitating (full line) and falling (dotted line) in state $a$ after
one, two, three and four sequences of four pulses as a function of
the frequency detuning
 $\delta \omega$ (kHz) ($\protect\omega_{1,2,3,4}:=\protect%
\omega_{1,2,3,4}^0+\protect\delta \protect\omega$) and for a
resonant interferometer duration $2T^0 \: \simeq \: 1.5 \: ms$.}
\label{fig:clock results}
\end{figure}

\begin{figure}[htbp]
\begin{center}
\includegraphics{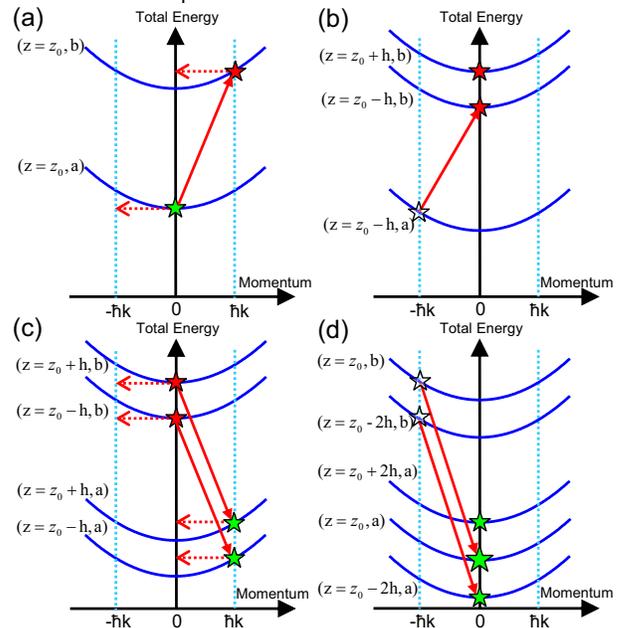}
\end{center}
\caption{(Color online)Motion of the atomic wave-packets in the
energy-momentum picture for the interferometer duration $2 T^0$.
Figs.~(a,b,c,d), associated with the 1st,2nd,3rd and 4th light
pulses respectively, show the packets present in coherent
superposition (full stars) immediately after - or transferred
(transparent stars) during - the considered pulse, whose effect is
represented by a full red arrow.} \label{fig:energy momentum}
\end{figure}

An analysis of the atomic motion in momentum space, sketched in the
energy-momentum diagram of Fig.~\ref{fig:energy momentum}, is
especially enlightening. In this picture, the total energy accounts
for the rest mass, the kinetic and the gravitational potential
energy. It is a parabolic function of the momentum. Each star stands
for a specific wave-packet, the motion of which between the light
pulses is represented by horizontal dashed arrows, in accordance
with energy conservation. For the duration $T:=T^0$, and for a
sufficiently coherent atomic sample, Fig.~\ref{fig:energy momentum}
reveals that the atomic motion in momentum is periodic and bounded
between two well-defined values associated with the photon recoil.
The momentum confinement is here provided by destructive
interferences which shut off the quantum channels going out of this
bounded momentum region. This remarkable property suggests an
analogy with an atomic resonator in momentum space. Following this
picture, we have computed the lower-state wave-function after $N$
resonant pulse sequences of duration $2T_0$, considering only the
vertical axis with no loss of generality. Each wave-packet ends up
at rest, and with a momentum dispersion $\Delta p_f$. Applying the
phase relation~\eqref{eq:basic condition} successively between the
multiple arms, one obtains: $\psi_a(p ,t_0+2 N T_0) = C_N e^{-\frac
{p^2} {\Delta p_f^2} } \sum_{Paths} e^{-i z_f p}$. $C_N$ is a
complex number, and the altitudes $z_f$ are the endpoints of the
resonant paths drawn on Fig.~\ref{fig:arches}, on which the sum is
performed. By labelling these paths with the instants of momentum
transfer, this sum appears up to a global phase as an effective
canonical partition function of $N$ independent particles, with
$Z_1=2 \cos^2 \left( k T p/2 m \right)$ the one-particle partition
function. This yields a wave-function of the form $\psi_a(p,t_0+ 2 N
T_0) =C'_N e^{i\phi(\mathbf{p},N)} e^{-\frac {p^2} {\Delta p_f^2} }
\cos^{2N} \left( p/p_{m} \right)$, with $p_m=2 m / k T_0$. As $N
\rightarrow +\infty$, multiple wave interferences thus yield an
exponential momentum localization, scaled by the momentum $p_m$,
around the rest value $p=0$. The diffusion in altitude observed in
the network of paths of Fig.~\ref{fig:arches} reflects a back-action
of this localization.

To summarize, we have proposed a space-time atomic sensor achieving
the levitation of an atomic sample through multiple wave
interference effects in a series of vertical $\pi/2$ light pulses.
The sensitivity of this levitation towards a double resonance
condition can be used to realize a frequency or an acceleration
measurement, with a sensitivity improving with the number of
interfering wave-packets. The sample needs to be cooled at a
nanoKelvin temperature in order to yield the desired interference
effects. At resonance, constructive multiple wave interferences then
maintain the full atomic population in suspension in spite of the
great number of non-levitating paths. For a sufficiently diluted
cloud, transverse confinement may be provided by the wave-front of
spherical light pulses. In this system, light shifts are due only to
a resonant light field and thus expected to be small. This proposal
opens promising perspectives for the development of cold atom
gravimeters~\cite{Chu99} and optical
clocks~\cite{Wilpers07,Trebst01,BestOpticalClocks,Katori03}. It may
also be turned into an atomic
gyrometer~\cite{BordeAtomInt89,Canuel06} by using additional
horizontal light pulses and exploiting the transverse wave-packet
motion.\\

\textit{Note added in proof:} This system may also work with
ultracold fermionic clouds. As the recent
experiment~\cite{Karski09}, our system implements a quantum random
walk~\cite{Aharonov03}, but here it involves a macroscopic number of
atoms propagating in free space.

%As shown, this setup is a passive atomic resonator~\cite{Wilkens93} in momentum space, it could nonetheless
%also be turned into an active system through matter-wave
%amplification by plunging the levitating wave-packets into a
%reservoir of source atoms.

The authors thank S.~Bize, P.~Bouyer, A.~Clairon, A.~Landragin,
Y.~LeCoq, P.~Lemonde, F.~Pereira, P. Wolf for stimulating
discussions, and A.~Landragin, S.~Walborn for manuscript reading and
suggestions. F.I. thanks N.~Zagury and L.~Davidovich for
hospitality. This work is supported by CNRS, DGA (Contract No
0860003) and Ecole Polytechnique.


\begin{thebibliography}{14}
\bibitem{footnote1} This
assumption does not induce any loss of generality: the following
discussion would also apply to any mode of the Hermite-Gauss
basis, and thus to any wave-function by linearity.
\bibitem{footnote3} In the configuration of Fig.~\ref{fig:arches}, the two pulse pairs are performed
successively, which maximizes the interferometric area for a fixed
interferometer duration; one could nonetheless also let a finite
time between them.
\end{thebibliography}
\end{document}